\documentclass[aps,prl,floats,twocolumn]{revtex4-1}

\usepackage{color}
\usepackage{graphicx}
\usepackage{graphics}
\usepackage{epsfig}
\usepackage{amssymb} 
\graphicspath{{./Fig/}}

\begin{document}


%
%
%
%

\title{Conductance of a photochromic molecular switch with graphene leads}


\author{C. Motta}
\email[]{c.motta10@campus.unimib.it}
\affiliation{Dipartimento di Scienza dei Materiali and Corimav Pirelli, Universit\`a di
Milano-Bicocca, via Cozzi 53, 20125 Milano (Italy)}

\author{M.I. Trioni}
\affiliation{CNR - National Research
Council of Italy, ISTM, Via Golgi 19, 20133 Milano, Italy}

\author{G.P. Brivio}
\affiliation{ETSF, CNISM and Dipartimento di Scienza dei Materiali, Universit\`a di 
Milano-Bicocca, via Cozzi 53, 20125 Milano (Italy)}

\author{K.L. Sebastian}
\affiliation{Department of Inorganic and Physical Chemistry, Indian
Institute of Science, Bangalore (India)}

\date{\today}

\begin{abstract} 
We report a full self-consistent \textit{ab initio} calculation of the conductance of a diarylethene-based 
molecular switch with two graphene electrodes. Our result show the contributions of the resonant states 
of the molecule, of the electrode density of states, and of graphene unique features such as edge states.
The conductivities are found to be significantly different for the two photochromic isomers at zero 
and finite applied bias. Further we point out the possibility of causing the switching by the application 
of a large potential difference between the two electrodes.
\end{abstract}
\maketitle


Organic electronics has rapidly grown into a
fundamental field whose potential is still to be fully developed and exploited.  Research
focuses on novel functional organic materials and existing applications already comprise,
among others, nanoscale electronic devices such as thin film transistors
and diodes, solar cells, integrated circuits on flexible substrates, carbon nanotube
field effect transistors and molecular switches.  A promising class of
organic molecular-scale photoswitching units is provided by photochromic molecules which are
able to switch between two chemical structures when irradiated by light of appropriate
wavelengths.  Such states correspond to photochemically interconvertible isomers
\cite{irie02}.

A recently developed class of  molecules, named diarylethenes,
displays a photochromic,  thermally irreversible, reactivity \cite{Irie_2010,
Nakamura_Yokojima_Uchida_Tsujioka_Goldberg_Murakami_Shinoda_Mikami_Kobayashi_Kobatake_2008}.  Most of them
exhibit light induced switching both in solution and in single
crystals \cite{Irie_2010,sci-irie,nat-irie} by changing
reversibly in aromaticity during the closed-open (open-closed) configuration transition, determined by
visible and ultraviolet light, respectively \cite{gnt1}.  This property  raised much interest and 
stimulated research on diarylethene derivatives
as suitable photochromic molecular switches \cite{katsonis,vd10}. Investigations of diarylenthene junctions with gold
electrodes showed that conductivity increases by two orders of magnitude in the closed isomer
compared to that of the open one \cite{he2005}.
Nonetheless several diarylethene derivatives on gold leads only display photochromic switching from the
closed to the open state after irradiation with visible light but not the reverse process
by UV \cite{duli03,vdrm06} owing to the fast
quenching of the photo-excited hole of the HOMO state of the open isomer 
into gold \cite{li2004}. By connecting the metal anchoring sulfur atom 
 by a phenyl group spacer reversible light-induced switching
resulted both on Au(111) \cite{Kats06} and gold nanoparticles \cite{kude06}. 
Photochromic diarylethene derivatives are  also
investigated on single walled carbon nanotube (SWCNT) as perspective single
molecule devices with a more accurately defined contacts to the electrodes
\cite{Guo_Small_Klare_Wang_Purewal_Tam_Hong_Caldwell_Huang_Obrien_2006}.
 For such system it was found that switching from insulating to conducting, i.e.  from open to closed
molecular configuration occurs, but not the reverse,  unlike in the case of Au
\cite{sper-graphene}. 
The timely character of this topic is confirmed by the wealth of theoretical studies on the conductance of
photochromic switches 
 both with gold \cite{li2004,kkon05,jcp2007,jpc2007,Nakamura_Yokojima_Uchida_Tsujioka_Goldberg_Murakami_Shinoda_Mikami_Kobayashi_Kobatake_2008,
su09,Odell_Sanvito_2010} and SWCNT electrodes \cite{zhaoSS09}.


In this paper we investigate the conductance of a diaylethene switch sandwiched between
two semiinfinite graphene electrodes by the Landauer formalism recast in terms of the non-equilibrium Green's function
(NEGF)   within density functional theory (DFT) \cite{datta}.
In fact, the two-dimensional (2D) form of carbon has just revealed several intriguing fundamental properties
and huge potential in nanoelectronics.
In particular, high chemical stability, low resistivity and
mechanical strength suggest  graphene,  a more cost-effective material than carbon
nanotubes, as an alternative component for
electrodes in electronic devices \cite{pang209}.
Furthermore in graphene electrons can travel ballistically up to a $\mu$m \cite{jian08},  and
electron-phonon scattering intensity is very weak \cite{moro08}.  This implies a
carrier mobility larger than that of the inorganic semiconductor with the largest mobility,
i.e. InSb, and that of carbon nanotubes.  If defects and impurities are eliminated, such
mobility is about $2\cdot 10^{5}$~cm$^2$/Vs \cite{jian08}, but in practice graphene sheets have
irregular shapes and contain impurities and defects.  To overcome this difficulty we point out a promising
bottom up experimental technique capable to synthetize graphene starting
from small corenene blocks and to build
well defined molecule electrode contacts \cite{diez10}.
Transparent, conductive, and ultrathin graphene films, have been demonstrated to be alternative metal-oxide electrodes for
solid-state dye-sensitized solar cells \cite{wang08}.

Our calculations show that at zero bias the peaks in conductance of the diarylethene moiety attached to two
graphene electrodes reflect not only the molecule electronic states broadened into resonances by the
coupling to the leads, but also the features of graphene edge states. 
Near the Fermi energy ($E_{\rm F}$) the conductance of the closed isomer is much lower in the energy interval between the HOMO and the LUMO 
resonances of the molecule,
following the linear dependence of the electrode density of states (DOS) proportional to $|E - E_{\rm F}|$. In the same range
the conductance of the open isomer is about zero.
If we apply a finite bias we observe that conductance is allowed within different energy windows
for the open and closed isomers.

\begin{figure} \includegraphics[width=0.48\textwidth]{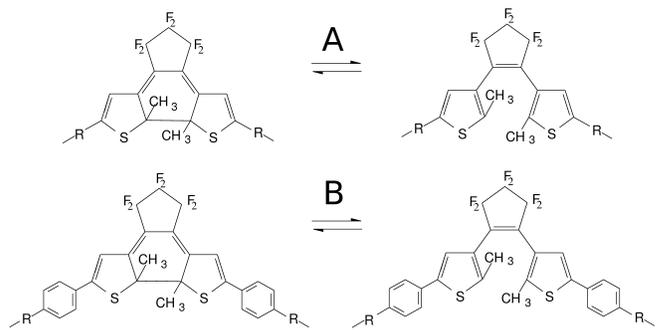}
\caption{\label{fig:isomers} Structures of the closed and open forms of the two diarylethene
derivatives studied in this work. Here, R is the functional group that links the molecule to
 graphene, which is $-$CH$_2-$ in the first case (A1 and B1) and $-$CO$-$NH$-$ (amide
linking) in the second case (A2 and B2)} \end{figure}

 We studied the open and closed structures of two  different diarylethene molecules, shown in  Fig.~\ref{fig:isomers}.  The central dithienylethene part is the
same for each molecule, and it is attached to the electrodes through $-$R$-$ with  R$=$CH$_2$ (we call this A1) and R$=$CO$-$NH
(referred to as A2).  We also considered longer isomers, which have two phenyl linkers and are
connected in the same fashion, namely B1 and B2. 
We started with optimized structures of the molecules with the two R groups connected to H atoms.  
Then the two H atoms were removed and the molecules were embedded between two graphene sheets. 
The graphene electrodes were taken to have zig-zag edges with the valencies of the edge C atoms satisfied by H atoms.  
We did this because C atoms at zig-zag terminations display stronger affinity towards radicals, and hence form stronger bonds with H atoms.   
Graphene with zig-zag edges have unique edge 
 state near the Fermi level \cite{jian07}. 
Further, we also optimized the energy by varying the electrode-electrode distance.   
We remark that in the optimized geometry, the plane of the molecule is approximately perpendicular to the plane of the two graphene sheets. 
This minimizes the coupling of $\pi$ states of the molecule with those of graphene, which can be advantageous in certain cases.  
In fact, it would be interesting to tune the orientation of the molecular plane with respect to that of graphene, by linking the molecule to more 
than one carbon of the graphene.   
In this way, one could adjust the hybridization of the molecular levels
by changing the relative orientation of the molecule with respect to the graphene plane.
The system may thus be thought of comprising of three separate parts:  the left and right semi-infinite graphene leads, and
an extended-molecule region.  
For each electrode, six layers of carbon atoms are included in the extended molecule region in order
to screen the perturbation from the central region.  
With  this, the final geometry with the contacts  is relaxed again.  
The graphene electrodes are constructed so that the
periodic replicas of the contact region are separated by four unit cells (i.e. 7.436~\AA),
and the interaction among them is negligible.  Our electronic structure calculations are
carried out using the DFT self-consistent pseudopotential method implemented in
SIESTA package \cite{SP01}.  The
exchange-correlation effect and electron-ion interaction is described by the
Perdew-Burke-Ernzerhof generalized gradient approximation and the norm-conserving
pseudopotential in the fully nonlocal form, respectively.  A double-$\zeta$ plus polarization
basis set is used to describe the localized atomic orbitals and an energy cutoff for real
space mesh size is set to be 200~Ry.  The ionic positions of carbon atoms are the ideal
positions of a previously relaxed graphene sheet, while the hydrogen positions are relaxed
for each junction. We performed all relaxations with a force tolerance of 0.04~eV/\AA.  For
the self-consistent calculations, 12 K-points are used in the transversal direction (that of
the electrode's edge), while for the calculations of the transmission function 60
K-points are taken into account.  The vacuum portions between two neighboring layers are set to be 15~\AA.
The transmission function of the system is calculated by the TranSiesta code \cite{bran02} 
which combines the NEGF technique with DFT. The results are similar for
the four molecules shown in Fig.~\ref{fig:isomers} and we choose to discuss those of B1 in depth.

%
%
%
%
%
%
%
%
%
%
\begin{figure} \includegraphics[angle=-90,width=0.49\textwidth]{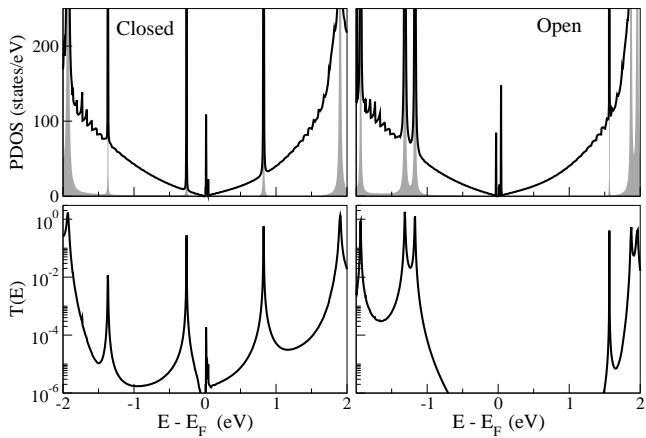}
\caption{\label{fig:dos-transm} PDOS and transmission function for the closed (left) and open (right)
isomers of B1 diarylethene. Upper panel: DOS (line) and PDOS (shaded area)
on the molecular region. Lower panels: zero-bias transmission functions.} \end{figure}
\begin{figure} \includegraphics[width=0.45\textwidth]{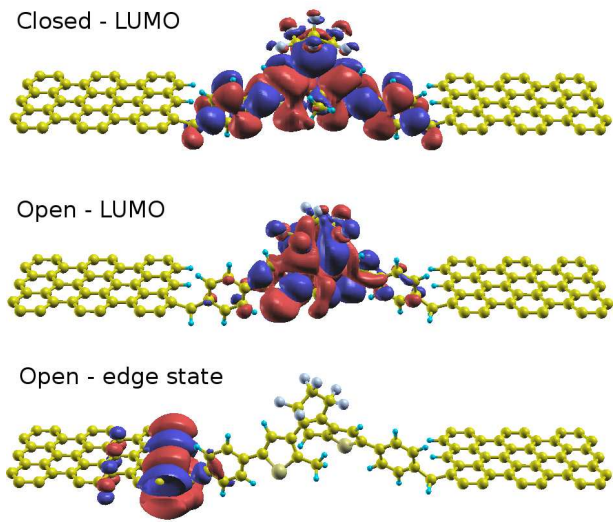}
\caption{\label{fig:autofunz} Real part of some representative wavefunctions of the junction with molecule B1: 
the LUMO of the closed isomer (upper panel), the LUMO of the open isomer (middle panel), 
and one edge state for the open isomer (lower panel).} \end{figure}
In the upper panels of Fig.~\ref{fig:dos-transm} we present the  
total DOS (solid line) of the extended system and the projected density of states 
(PDOS) (shaded area) onto the molecular region, as function of the electron
energy, the left and the right one for the closed and open isomer, respectively.
The transmission function dependence on the
injected electron energy is shown in logarithmic scale in the lower panels of Fig.~\ref{fig:dos-transm}.
The total DOS is reminescent of that of the
electrodes, since near $E_{\rm F}$  it displays a linear dispersion following that
of an ideal, infinite graphene sheet.  
Of course, very close to  $E_{\rm F}$ the dispersion is not perfectly linear due to the presence of a vacuum region
between two graphene sheets.
Furthemore we remark the presence of two low peaks very near $E_{\rm F}$,
one slightly above and the other just below it.  They are states mainly localized at the molecule-electrode
interface whose origin stems from the zig-zag termination edge states of the two graphene sheets
 interacting with the orbitals of the molecule in the coupled system.
We point out that, as a consequence of the perturbation due to the junction,
 such states lie within the continuum graphene spectrum and 
they become resonant conducting states, which do not exist in SWCNT \cite{diez10}.
Apart from the edge states,
all peaks in the PDOS refer to molecular orbitals broadened by the 
interaction with the electrodes and a corresponding feature at the same energy
appears in the transmission function. We concentrate on the two main peaks of conductance for each isomer that 
appear on the left and on the right side of $E_{\rm F}$. They are due
to the hybridization of the HOMO and LUMO states of the
molecule with the graphene substrate.  
For the closed isomer such states are closer to $E_{\rm F}$, and
their intensity is lower, if compared with that of
other molecular states, since the DOS of the underlying graphene sheets tends to zero 
as the energy approaches $E_{\rm F}$.
However, considering small enough energy
deviations from $E_{\rm F}$, the transmission function is always larger for the closed isomers than
for the open ones.  The lower conductance of the open isomers is due to their distorted
geometry: the breaking of the central C$-$C bond due to the closed-open transition affects the
$\pi$-conjugation of the structure  reducing it \cite{maror2000}.  Hence, the
electrons are less delocalized along the molecular backbone reducing  transmission.
In Fig.~\ref{fig:autofunz} we display the real part of the wavefunction for three representative
states of the system, namely the LUMO of the closed isomer, the LUMO of the open isomer, and one 
edge state of the open molecule. We observe that the former LUMO is more widely spread  than 
the latter one, reflecting its more extended $\pi$ conjugation. The plotted edge state shows
a rapid decay inside the molecular region. The other edge state just mirrors it on the opposite side of the junction.
Comparing the transmission functions of diarylethene with different connecting
groups (B1 and B2), we noticed that the presence of the amide end-groups in the latter
molecule produces a slight
shift of the peaks toward higher energies. This results in a better alignment of the HOMO
of the closed isomer with $E_{\rm F}$ so, once a voltage is applied, 
a larger current may be obtained in this case.  
But, as the HOMO state is shifted nearer to $E_{\rm F}$, its conductance  
intensity is lowered because the DOS's of the electrodes diminish.
Apart from this shift, the structure of the transmission function is not appreciably
affected by the change of connecting groups.

\begin{figure}
\includegraphics[width=0.34\textwidth,angle=-90]{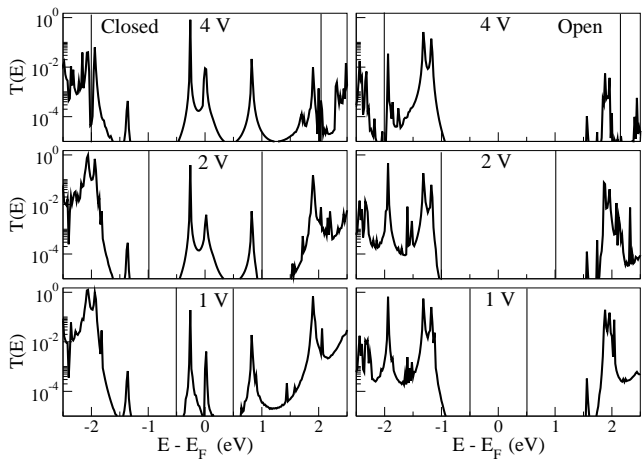}
\caption{\label{fig:bias} Transmission function of the closed (left) and open (right)
 isomers of the B1 diarylethene derivative at different applied biases (from bottom to top: 1V, 2V, 4V).
The bias window is delimited by two vertical lines.} 
\end{figure}
%
We now discuss the changes in the conductance for molecule B1 when the system is driven out of equilibrium 
by the application of a bias $V$.
In Fig.~\ref{fig:bias} the transmission functions are reported for the closed (left panel) and open
isomer (right panel) by increasing the voltage from the bottom pictures. For the closed
molecule a bias of 1~V allows one to include the contribution of the HOMO 
in the admissible energy window (vertical lines), while a bias of 2~V is needed
to contain the LUMO. 
We also observe two features from the edge resonances near $E_{\rm F}$.
In fact, as a finite bias is applied, edge states play a role, though small, in the transmission curve.
A non negligible conductance asks for a bias $V$ = 4 V for the open isomer. 
In both cases the HOMO-LUMO energy difference is unaffected by the applied voltage.

In solution, diarylethene based molecules can be
switched between the two states by photochemical means.  When used as a bridge between two Au
electrodes, the switching occurs, usually in one direction only, due to quenching by the
electrodes.  For CNT electrodes too, the switching from open to
closed isomers occurs through light, but the reverse does not \cite{sper-graphene}.  
We suggest the use of graphene as the electrode as (1) it has a lower density of states near the Fermi level 
in comparison with Au leading to lesser probability for quenching and (2) the orientation of the molecule
 with respect to the graphene sheet can be changed chemically, tuning the interaction between the two components.   
In the polymeric state, the change from closed state to the open state can be induced
not only by light, but also by electron/hole injection obtained by the application of a potential
difference
\cite{Nakamura_Yokojima_Uchida_Tsujioka_Goldberg_Murakami_Shinoda_Mikami_Kobayashi_Kobatake_2008}.
Consequently one may anticipate that in the case of diarylethene molecules
connected to graphene, the switching from open to closed may be induced by light and the
reverse by the application of a large potential difference.  
Looking at the positions of the LUMO/HOMO orbitals energies in the transmission curves
given in Fig.~\ref{fig:bias}, it seems possible to populate/depopulate them to a sizeable
extent by the application of about $2$ and $4$ volts
for the closed and open configurations, respectively. Thus one has the possibility of inducing
switching in both directions, by the application of such a potential difference for a short
period of time, and hence light may be avoided altogether.

In conclusion we have calculated the conductance of the two isomers of diarylethene-based molecules and
analyzed the role of the graphene leads. We have pointed out the advantages of using graphene, namely
a reduced quenching of the photoexcited state compared to gold electrodes and the possibility to tune the resonant 
HOMO and LUMO hybridization by orientating the molecule with respect to the leads plane.
We also stress that, in the case of graphene, 
the switch can be operated electrically by the application of short pulses, making it potentially very useful, 
in comparison with  other electrodes. 

\section{Aknowledgment} C.M. and K.L.S. thank CARIPLO Foundation for its support within the
PCAM European Doctoral Programme.




\providecommand{\noopsort}[1]{}\providecommand{\singleletter}[1]{#1}%
%

\end{document}